# Review of strategies for the reuse of N95 and N99 respiratory masks during the COVID-19 pandemic

Hans Joachim Schöpe and Miriam Klopotek
Institute for Applied Physics, University of Tübingen, 72076 Tübingen, Germany
Corresponding author e-mail: hans-joachim.schoepe@uni-tuebingen.de

The COVID-19 pandemic presents a strain of unprecedented scale on health systems around the world. We are currently witnessing catastrophic bottlenecks in the supply of essential medical products as a result of major disturbances in the global economic system. In addition to medical ventilators, personal protective equipment is in extreme shortage and, hence, currently in the spotlight of the media. N95 and N99 (European FFP2 and FFP3) respiratory masks play an especial role. In order to reliably protect medical personnel, and thus to contain the spread of the pandemic, it is essential to provide at least N95 respiratory masks (FFRs). The WHO estimates that about 89 million FFRs are needed each month to combat the COVID-19 pandemic – with such estimates rising [1]. Meeting this demand implies not only a ~40% increase in the global production of FFRs, but also ultra-rapid delivery of the masks to hospitals, given the current situation. Both are rather improbable. One way to relax this precarious situation is to reuse existing masks several times. Of course, they would thereunto have to be decontaminated after each use. This procedure is not at all ubiquitous in the medical system, as FFRs are intended for single use by manufacturers. However, in recent years the scientific literature has laid out various possibilities to decontaminate FFRs and reuse them several times [2,3,4,5,6,7]. All these methods must meet the following requirements: the masks must be decontaminated, the effectiveness of the mask must not be damaged, the user must not be endangered in any other way (e.g. poisoning) and the masks should be ready for reuse quickly. In addition, the methods should avoid the usage of disinfectants, as these are in short supply in the current situation.

Some of the methods examined in the above-cited articles meet these criteria. In the following we classify these methods according to their mode of operation and the availability of the corresponding equipment as follows:
1. UVc-radiation (disinfection on an industrial scale using professional UV-sterilization systems)
2. Heat treatment with moist heat (heat treatment by means of ovens or dryers)
3. Microwave-generated steam (microwave ovens, on-site application with minimum effort)

The SARS-CoV-2 virus does not differ dramatically from other viruses in its chemical and physical resistance. First investigations show that SARS-CoV-2 can be eliminated in a few minutes with the usual disinfectants, UVc irradiation, or temperatures above 56°C [8,9,10,11]. The results for the SARS-CoV-2 virus are comparable with those of the more recent pathogens of the last years (SARS-CoV-1, H1N1, H5N1) [4,6,7,12,13,14,15]. The table summarizes some of the existing (preliminary) results and compares them directly with data of the influenza-A pathogen H1N1. The disinfectants used included, among others, ethanol (75%), hand soap solution (2%), Clorox (10%), acetone and household bleach [8,9].

| *Disinfection method* | *SARS-CoV-2* | *H1N1* |
| --- | --- | --- |
| Disinfectant | less than 5 min | less than 2 min |
| Temperature | ca. 15 min at 56°C<br>ca. 1 min at 70°C | 5 min at 70°C |
| UVc-radiation (256nm) | A few minutes[1]<br>required dose: ~ 1 J/cm$^2$ | A few minutes*<br>required dose: ~ 1 J/cm$^2$ |

Table: Times required to kill the viruses on surfaces using the listed methods.

It is therefore clear that, in principle, the methods described in the literature for FFR decontamination of H1N1, H5N1 or SARS-CoV-1 can be used successfully in the COVID-19 pandemic with the aim of re-using FFRs several times. In the following, three different methods are introduced briefly and further procedures are discussed.

1) UV-disinfection

UVc light has been utilized for the sterilization of liquids, gases and solid surfaces since the middle of the 20th century. It is used widely in the life sciences, but also intensively in the food industry and for drinking water purification [16]. To decontaminate FFRs, the masks must be illuminated on both sides with a sufficiently high radiation dose by means of one or more UVc lamp(s). In the simplest version (e.g. Ref. 4), an 80W UVc lamp with 256nm wavelength of about 1m length is used for illumination at a distance of 25cm. The investigations showed that influenza pathogens (H1N1 and H5N1) are removed up to 99.9% when using a dose of about 1J/cm$^2$. Using this simple set-up, this dose is reached within a few minutes. The required decontamination dose for SARS-CoV-2 does not differ significantly from that for the influenza pathogens [10]; thus, UV disinfection can be used in a fully analogous way during the COVID-19 pandemic. The decontamination time can be reduced massively when using several UVc lamps with higher output power simultaneously, as in the food industry, and decontamination can be processed in "production line mode" [7]. The UVc disinfection units located in hospitals could be put to use for this purpose: It would be an on-the-spot solution. Research laboratories often have such facilities, too. Furthermore, it would be highly desirable to use the existing high-throughput UVc disinfection systems from other industrial branches (e.g. sterilization of cans or yogurt cups) for FFR disinfection. To this end, used masks would have to be collected and delivered to an appropriate company in order to disinfect them on a large scale. The UVc procedure should disinfect the FFRs from nearly all common bacteria and for a large number of common viruses [16].

---

[1] The irradiation time required to achieve a dose of 1J/cm$^2$ depends on the output power of the UVc light source and the geometry of the set-up.

2) Warm moist heat

The germicidal effect of heat is widely known – the addition of moisture accelerates this. Steam condenses on colder objects until these have reached the temperature of the steam. To decontaminate FFRs, the masks must be exposed to a warm, humid atmosphere for a sufficiently long time. Several studies have shown that influenza pathogens (H1N1 and H5N1) can be removed up to 99.9% using a rather simple procedure [3,4,6]. In one of the publications, a sealable 6l vessel was filled with 1l of tap water and heated to 65°C in an oven. The container was then removed from the oven and a respirator mask was positioned above the water surface using a rack. The container was then closed again and placed back into the oven for a 20-minute treatment. The samples were then air-dried overnight. This method can also be used in the COVID-19 pandemic to remove SARS-CoV-2. Upscaling the method is straightforward. To shorten the drying time, placing the treated masks in a drying cabinet is useful.

Comment: There is a risk of contaminating the room in which the oven is placed, especially if the oven has a ventilator when turned on. Hence it is absolutely necessary that the container within which the mask is placed is sealed!

3) Microwave-generated steam

This method is related to the one above, as it includes steam used for disinfection. Two different procedures are described in the literature. In the first [3,4,6], two small plastic containers with perforated surfaces (e.g. storage boxes for pipette tips) are filled with 50ml tap water and placed side-by-side in a microwave oven. The FFR is placed face-down on the pipette tip boxes and is then irradiated using a power of 1250 W for 2 minutes. After treatment, the pipette tip boxes are refilled with fresh tap water and the next FFR is decontaminated. In the second procedure [5], the FFR is placed in a 2l microwave steam bag, pouring 60ml tap water into it. These microwave bags are commonly used for steam cooking as well as for disinfecting infant pacifiers or drinking bottles. The bag is sealed and then irradiated with 1100 W for 90 seconds in a microwave oven. It has been shown that these simple procedures can remove 99.9% of influenza pathogens. Limitations: These methods are not easily applicable to **all** FFRs. If metal parts are installed in the mask, then there is a risk that the mask could be damaged due to strong local heat generation. It would be possible to remove the metal part (nasal clamp) temporarily given this can be done easily.

The methods listed have been known for rather long and have been published in scientific journals about ten years ago. They offer the possibility to decontaminate the currently most urgently needed FFRs, and thus to use them repeatedly. Furthermore, modifications are possible for the listed processes. The following example demonstrates a modification that is realizable in a timely manner and uses devices available on the market without major, additional developments: For disinfection with moist heat, standard heat pump tumble dryers for laundry could also be used, as many models work at a temperature of approximately $65\pm5$°C. The FFRs are first moistened with tap water and then dried in the tumble dryer. A drying temperature above 56°C ensures that SARS-CoV-2 viruses are removed, but the temperature is low enough to avoid damage to the FFRs. This would be the case for vented or condenser tumble dryers, as their working temperature is around 100°C. Professional laundries – some of which are affiliated with hospitals – may have suitable dryers and could carry out the disinfection of a larger quantity of FFRs directly on site.

A recent technical bulletin from the FFR manufacturer 3M [17] indicates that treatment similar to that described above may compromise the mask fitting: The nosefoam may deform and the elastic straps may lose elasticity. Similar observations were reported in the investigations cited above, where a strong dependence on the particular mask model, however, was indicated. There is agreement among the investigations that the filtering function of the FFRs is not compromised at all after such treatments. To restore the fitting function, the nosefoam and straps could be replaced after the disinfection procedure.

In this paper, we reviewed strategies for disinfecting FFP2, FFP3, N95 and N99 respiratory masks (FFRs) for their reuse based on the existing scientific literature. These methods are intended as emergency solutions only –  for situations where there is a critical shortage of the single-use respiratory masks. The methods described here can be done on-site within e.g. a hospital or a medical office, given there is sufficient equipment available, or up-scaled with the help of existing infrastructure in the industry. In particular, two of the described the methods (warm moist heat and microwave-generated steam) require essentially standard kitchen appliances. All three methods were shown to be successful for disinfecting masks from pathogens. The method using UVc radiation seems to offer a very robust and very effective solution, but requires more technical skills to calculate the correct irradiation time to achieve the required dose.

The strategies reviewed in this work could help ease the current precarious situation globally – and thus save lives.

**Disclaimer**: The methods and procedures outlined here are intended for "self-help" in emergency situations only and are not certified in any way. Anyone choosing to execute on these methods does so at his/her own risk.

It would be highly desirable for a governing instance to validate or officially recommend any of these methods for the time of the pandemic.

---